\documentclass[twocolumn]{aastex631}

\usepackage{bm}

\shorttitle{The role of meridional flow in the generation of solar/stellar magnetic fields and cycles}
\shortauthors{Vashishth \& Karak}
\newcommand{\Fig}[1]{Figure~\ref{#1}}

\newcommand{\Eq}[1]{Equation~(\ref{#1})}

\newcommand{\Sec}[1]{Section~\ref{#1}}

\def\bl{Babcock--Leighton}
\def\mc{meridional circulation}
\def\mf{meridional flow}

\def\Rs{R_{s}}
\newcommand{\mps}{m~s$^{-1}$}

\newcommand{\etas}{\eta_{\mathrm{S}}}
\newcommand{\etaCZ}{\eta_{\mathrm{CZ}}}
\newcommand{\etaRZ}{\eta_{\mathrm{RZ}}}
\newcommand{\rBCZ}{r_{\mathrm{BCZ}}}
\newcommand{\rsurf}{r_{\mathrm{surf}}}

\begin{document}

%\title{Exploring the role of meridional circulation in the solar and stellar dynamos}
\title{The role of meridional flow in the generation of solar/stellar magnetic fields and cycles}

\correspondingauthor{Bidya Binay Karak}
\email{karak.phy@iitbhu.ac.in}

%[0000-0002-2296-9141]
\author{Vindya Vashishth}
\affiliation{Department of Physics,  
Indian Institute of Technology (BHU), 
Varanasi 221005, India}
%[0000-0002-8883-3562]
\author{Bidya Binay Karak}
\affiliation{Department of Physics,  
Indian Institute of Technology (BHU), 
Varanasi 221005, India}

\begin{abstract}

Meridional flow is crucial in generating the solar poloidal magnetic field by facilitating the poleward transport of 
the field from the decayed Bipolar Magnetic Regions (BMRs).
As the meridional circulation changes with the stellar rotation rate, the properties of stellar magnetic cycles are expected to be influenced by this flow.
In this study, we explore the role of meridional flow in generating magnetic fields in Sun and sun-like stars using 
STABLE [Surface flux Transport And Babcock–LEighton] dynamo model.
We find that a moderate meridional flow increases the polar field by efficiently driving the trailing polarity flux toward the pole, while a strong flow tends to transport both polarities of BMRs poleward, potentially reducing the polar field. Our findings are in perfect agreement with what one can expect from the surface flux transport model.
%We find that STABLE captures the response of meridional flow more accurately as compared to the traditional 2D dynamo models using $\alpha$ parameterization for the Babcock--Leighton process.
%In particular, our findings align with Surface Flux Transport (SFT) models, showing that a moderate meridional flow increases the polar field by efficiently driving the trailing polarity flux toward the pole, while a strong flow tends to transport both polarities of BMRs poleward, potentially reducing the polar field.
Similarly, the toroidal field initially increases with moderate flow speeds and then decreases after a certain value. This trend is due to the competitive effects of shearing and diffusion.
Furthermore, our study highlights the impact of meridional flow on the cycle strength and duration in stellar cycles. 
By including the meridional flow from a mean-field hydrodynamics model in STABLE, we show that the magnetic field 
strength initially increases with the stellar rotation rate and then declines in rapidly rotating stars, offering an explanation of the observed variation of stellar magnetic field with rotation rate.
\end{abstract}


\section{Introduction} 
\label{sec:intro}
%General overview
After differential rotation, the second important axisymmetric flow inside the solar convection zone (CZ) is the \mc, which is a weak flow from equator to pole in the upper CZ and from pole to equator in the deeper CZ \citep{Kit16, Hanasoge22}. This flow has been observed on the surface for many years by tracking various tracers \citep[e.g.,][]{Makarov83, MS89}. 
%Depending on the measurement techniques, latitude band, and temporal window, an equatorward speed varying from about 10 to 30 \mps\ is detected near the surface. 
%While a mass conservation principle implies a return flow of a few \mps, due to difficulties of the measurement in the deep CZ, it is still not confirmed about the depth, speed, number of cells, and latitudinal profile of the return flow; however, see \citet{RA15, Gizon20}, for the recent developments in this field.

%Summary of theoretical studies (MF and global MHD modelling)\\
Hydrodynamic mean-field models---employing the mean-field versions of the equation of plasma flows, heat transport, and equation of states---have been employed to model the large-scale flows in the sun and stars in which the \mc\ is a natural consequence of the angular momentum balance \citep{Brand92, KR95, KS01, Remple05,  HY11}. 
Particularly, the solar differential rotation produced in the model of \citet{KO11b} remarkably matches with the helioseismic observations \citep[e.g.,][]{Schou98}. Further, the model agrees with the observations of surface differential rotation of rapidly rotating stars \citep{barnes05}. 
The \mf\ in the model of \citet{KO11b} consists of single-cell circulation with a surface flow speed of about 15 \mps\ and return flow of 5 \mps. With the increase of the rotation rate of the star, the flow becomes weaker (it is strong only near the radial boundaries). On the other hand, the global convection simulations also produce \mf\ in the solar and stellar CZs \citep[e.g.,][]{FM15}. The flows from these simulations are quite complicated, usually consisting of multiple cells and some signatures of poleward flow near the surface and return flow near the base of CZ \citep[see, e.g., Fig. 8 top right of][]{Kar15}. As the global convection simulations are still exploratory and struggling to reproduce the fundamental observations of solar and stellar CZs (e.g., convection speed at large length scale), we must interpret their results cautiously \citep{KMB18b, Kapy23rev}. 

%Importance in MC in generation of polar field\\
Meridional circulation plays an important role in the generation of the poloidal magnetic field near the surface. It is this flow that pushes the field towards the poles from the low latitude, where the poloidal field is generated through the decay and dispersal of the tilted bipolar magnetic regions (BMRs). Without this flow, the field from the decayed BMRs might largely gets cancelled, and a negligible amount of polar field is expected to be produced. The evolution of the surface magnetic field clearly imprints a poleward transport due to meridional flow \citep[in fact, historically, this feature has been the basis for the measurements of \mf\ on the surface by some authors, e.g.,][]{HL81, Topka82}. 
However, a strong \mf\ can cause both polarities of a BMR to move towards the pole, enabling inefficient cross-equatorial cancellation. Thus, we expect a non-monotonous variation of the polar field strength with the increase of the flow speed. Surface flux transport (SFT) simulations of \citet{Bau04}, by including the meridional flow, differential rotation, turbulent diffusion, and BMRs source, showed an increasing dependence of the polar field at low meridional flow speed and decreasing trend at high flow (also see \citet{Jiang10, UH14b} for the demonstration of the polar field variation due to fluctuations in meridional circulation).  
So far, no dynamo model has been able to reproduce this trend completely. 
By including double ring as the replacement of Babcock--Leighton $\alpha$ effect, \citet{munoz10} showed that the polar field decreases with the increase of meridional flow speed. However, their study was not performed at low meridional flow speed, and thus, the full spectrum of the polar field variation has not been reproduced so far. One motivation of our work is to show this dependence using a three-dimensional (3D) dynamo model. 

%Importance in MC in solar stellar dynamos\\
In addition to the generation of a poloidal field in the Sun, the meridional circulation plays a vital role in determining the features of the solar cycle via the dynamo process \citep{CD00, Cha04, Kar10, CK12, Haz15, SN2016, Chou21, Hazra23}. 
In flux transport dynamo models \citep{Kar14a, Cha20, K2023}, it is the flow that helps in transporting the polar field from the surface to the deeper CZ at high latitudes, where the differential rotation stretches the field to produce a toroidal one. At the base of CZ, the toroidal field is transported to the low latitude from where it gives rise to BMR eruptions. Thus, a shorter cycle period is expected at faster circulation speed. Indeed, previous axisymmetric kinematic flux transport dynamo models found an almost inverse relation for the cycle period with the meridional flow speed, namely, $P_{\rm cyc} \propto v_0^{-0.89}$ \citep{DC99,YNM08} or $P_{\rm cyc} \propto v_0^{-0.7}$ at high diffusivity regime \citep{Kar10}, where $P_{\rm cyc}$ is the cycle period and $v_0$ is the flow speed. 

Meridional flow also affects the strength of the toroidal magnetic field. The axisymmetric flux transport dynamo model showed that in the low diffusivity regime, decreasing meridional flow causes an increase in the field strength by giving the shear more time to induct a toroidal field. In the high diffusivity regime, this effect is overpowered by an opposing effect---weaker flow speed gives more time for the diffusion to diffuse the field, which eventually decreases the field \citep{YNM08}; also see \citet{KC10, KC11} for an application of this idea in modeling the Waldmeier Effect.

%Motivation and Aim\\
In the present study, we employ a 3D kinematic solar dynamo model STABLE \citep[Surface flux Transport And Babcock–LEighton;][]{MD14, MT16, KM17} to study the effect of meridional flow speed on the polar field, toroidal field and cycle duration. As in STABLE, the \bl\ process is realistically modeled, 
which can be operated as an SFT model in addition to the (normal) dynamo mode. 
In this work, we shall first check whether STABLE reproduces the expected variation of the polar field with the \mf\ as found in the SFT models. This is an essential step because any dynamo model should reproduce the basic features of the SFT model, which has proven successful in reproducing surface observations in great detail \citep{yeates23}. Next, we shall demonstrate how the cycle strength and duration vary with the \mf\ speed.
We shall show that the result from our comprehensive dynamo model is quite different than found in axisymmetric flux transport dynamo models. 
Our results will have an implication for stars whose cycles have different strengths and durations \citep[e.g,]{Baliu95, garg19}, and the meridional circulation speed changes with stellar rotation rates \citep{Brown08, KO11b, KO12b, Kar15, viv18}. Thus, our study will identify the meridional flow as a cause of the variations of cycle strength and duration of stellar cycles with the rotation rate. In \Sec{sec:models}, we present our model, while in \Sec{sec:results} we discuss our results. Finally, in \Sec{sec:conclusion}, we summarize our results and highlight the conclusion.

\section{Model} 
\label{sec:models}

\subsection{2D dynamo model with %$\alpha$ effect}
\bl\ source}
\label{sec:2d}
We initialize our study by using a 2D kinematic dynamo model \citep{CSD95,DC99} in which the Babcock--Leighton process is parameterized by an $\alpha$ term. In this model, the following equations for the axisymmetric magnetic field are evolved. 
\begin{equation}
\frac{\partial A}{\partial t} + \frac{1}{s}({\bm v_p}.\nabla)(s A)
= \eta_t \left( \nabla^2 - \frac{1}{s^2} \right) A + \alpha \overline{B}_{\rm BCZ},
\label{eq:pol}
\end{equation}
\begin{eqnarray}
\frac{\partial B}{\partial t}
+ \frac{1}{r} \left[ \frac{\partial}{\partial r}
(r v_r' B) + \frac{\partial}{\partial \theta}(v_{\theta} B) \right]
= \eta_t \left( \nabla^2 - \frac{1}{s^2} \right) B \nonumber \\
+\ s({\bm B_p} \cdot {\bf \nabla})\Omega + \frac{1}{r} \frac{d\eta_t}{dr} \frac{\partial} {\partial r} (rB),~~~
\label{eq:tor}
\end{eqnarray}\\
where 
${\bm B_p} = \nabla \times [ A(r, \theta) {\bf e}_{\phi}]$ is the poloidal component of the magnetic field,
$B(r, \theta)$ is the toroidal component, $s = r \sin \theta$, ${\bm v_p} =  v_r^\prime { \hat r} + v_{\theta} {\bf \hat \theta}  =  (v_r +\gamma_r) { \hat r} + v_{\theta} {\bf \hat \theta}$, which includes the meridional circulation ($v_r,v_\theta$) and the radial pumping ($\gamma$), $\eta_t$ is the effective turbulent diffusivity that incorporates the mixing effect of the small-scale convective flow, $\alpha$ is a parameter that captures the BL process in this 2D dynamo model and $\overline{B}_{\rm BCZ}$ is the average toroidal field in the thin layer from $0.69\Rs$ to $0.71\Rs$ ($\Rs$ is the radius of the sun), and $\Omega$ is the angular velocity.

For the meridional circulation, we consider a single-cell flow, poleward near the surface and equatorward at the base of CZ. 
For which, we consider a stream function $\psi$, in such a way that,
\begin{equation}
    \rho {\bm v_p} = \nabla \times [\psi(r,\theta){\hat \phi}],
\end{equation}
and, 
\begin{eqnarray}
\psi r \sin \theta = \psi_{\rm 0} (r-R_{\rm P}) \sin \left[\frac{\pi (r-R_{\rm P})}{R_{\rm s}-R_{\rm P}}\right]( 1 - e^{-\beta_1 \theta^{\epsilon}} ) \nonumber\\
\times (1 - e^{\beta_2(\theta - \pi/2)}) e^{-((r-r_0)/\Gamma)^2}.~~~
\end{eqnarray}\
Here, $\rho = C(\frac{R_s}{r}-0.95)^{3/2}$, is the nondimensional density stratification, $\beta_1$ \& $\beta_2$ are 1.5 \& 1.3, $\epsilon = 2.0$, $\Gamma = 3.47 \times 10^8$ m and $R_{\rm P} =0.69R_{\rm s}$. The amplitude of the meridional circulation at mid-latitude $v_{\rm 0}$ is obtained from the value of $\psi/C$.
%The profile follows the descriptions found in several prior studies, notably outlined in \citet[Equation (5)]{KC16}

For the differential rotation, we use an analytic function that aligns with observed helioseismic data and has been used in many previous publications; for example, see Equation (3) of \citet{MT16}.
%The ${\bm v_p}$ profile is obtained through an observationally-guided analytic formula as given in \citet[Equation (5)]{KC16}.
%and is given by $\bm \gamma = \gamma_r(r){\hat r }$ where, 
The profile of the radial magnetic pumping is 
%given by
%\begin{eqnarray}
% \gamma_r(r) 
 %= - \frac{\gamma_{CZ}}{2}
 %\left[1 + \mathrm{erf} \left(\frac{r - 0.725\Rs}{0.01\Rs}\right) \right]\nonumber \\
% - \frac{\gamma_S}{2} \left[1 + \mathrm{erf} \left(\frac{r - 0.9\Rs}{0.02\Rs}\right) \right].
%\label{pumping}
%\end{eqnarray}
%Here, $\gamma_{CZ}$ = $2$ \mps ~and  $\gamma_S$ = $20$ \mps. 
same as shown by \citet{Ca12, KC16}. This radial pumping suppresses the diffusion of the magnetic field through the surface and thus helps the model to operate at a high diffusivity value \citep{KM17}.

For $\eta_t$, we take it as a function of $r$ alone and has the following form:
\begin{eqnarray}
\eta_t(r) = \etaRZ + \frac{\etaCZ}{2}\left[1 + \mathrm{erf} \left(\frac{r - \rBCZ}
{d_1}\right) \right]\nonumber \\
+\frac{\etas}{2}\left[1 + \mathrm{erf} \left(\frac{r - \rsurf}
{d_2}\right) \right],
\label{eq:eta}
\end{eqnarray}\\
with $\rBCZ=0.715 \Rs$, $d_1=0.0125 \Rs$, $d_2=0.025 \Rs$, $\rsurf = 0.956 \Rs$,
$\etaRZ$, $\etaCZ$, and $\etas$ represent the diffusivities at the inner boundary, within CZ, and at the surface, respectively, having the values as $\etaRZ = 1.0 \times 10^9$ cm$^2$~s$^{-1}$, $\etaCZ = 1.5 \times 10^{12}$ cm$^2$~s$^{-1}$, and $\etas = 3 \times10^{12}$ cm$^2$ s$^{-1}$.

One motivation of our work is to understand the response of the poloidal field to the meridional flow of different level 
%at different levels of meridional flow 
and that we study by placing a single BMR in our dynamo model. However, in the present axisymmetric model, we mimic the generation of the poloidal field from a single BMR by localizing the poloidal source function in a narrow latitudinal band. 
Hence, we take the following profiles of $\alpha$.
\begin{equation}
\alpha = \frac{\alpha_0} {4} \left[1 + \mathrm{erf} \left(\frac{\theta - \theta_1}
{\Delta \theta}\right) \right] \nonumber
\left[1 - \mathrm{erf} \left(\frac{\theta - \theta_2}
{\Delta \theta}\right) \right],
\label{eq:alpha}
\end{equation}
where $\alpha_0 = 50 $ m~s$^{-1}$, $\Delta \theta = 2^{\circ}$, 
$\theta_1 = 16^{\circ}$, and $\theta_2 = 14^{\circ}$. 

No quenching in $\alpha$ or anywhere is included in this 2D model, thus making it fully linear.
%No quenching in $\alpha$ or anywhere is included, and thus, the model used in our study is fully linear.

\subsection{STABLE: 3D \bl\ dynamo model}
For most of our calculations, we use a 3D kinematic dynamo model,  STABLE \citep{MD14, MT16, KM17, HM16}, 
which solves the following induction equation in three dimensions within the solar CZ. 
\begin{equation}
\frac{\partial \bm B}{\partial t} = \nabla \times \left[  \bm V \times \bm B - \eta_t \nabla \times \bm B \right],
\label{eq:induction}
\end{equation}
where the large-scale velocity, $\bm V$ is represented as,
\begin{equation}
{\bm V} = {\bm v_p} + r \sin\theta~\Omega (r,\theta) { \hat \phi}.
\end{equation}
The profiles of $\bm {v_p}$, $\Omega$, and $\eta_t$, follow the same formulation as outlined in \Sec{sec:2d}. 

The major difference between the STABLE and 2D dynamo model is that in STABLE, we do not include $\alpha$ term for the poloidal source in \Eq{eq:induction}; rather, BMRs are deposited (based on the toroidal field at the base of CZ) on the surface, and then the decay of these spots produce a poloidal field. 

Further, in STABLE, the magnetic field of the sunspot is extrapolated below the surface using a linear potential field extrapolation of the surface fields. For additional details on the field distributions, we refer the readers to Section 3 of \cite{MD14}. 
As the STABLE includes explicit BMRs and captures their decay on the solar surface in the same way as in SFT models, we can operate STABLE in SFT mode by feeding the observed or synthetic BMR data.

\subsubsection{STABLE as SFT model with single BMR}
\label{sec:STABsingle}
Our primary motivation with STABLE is to explore the evolution of the polar field. To achieve this, we first use the STABLE model in the SFT mode and include a single BMR for the source of the poloidal field.

This BMR is placed at a certain 
latitude of the northern hemisphere and has the following properties: (i) 
Tilt as given by Joy's law i.e., Tilt = $30 ^\circ \sin \lambda$ (where $\lambda$ is latitude),
(ii) Flux in each pole of BMR ($\Phi$) = $10^{22}$~Mx (within the observational range, as mentioned in \citet{Anu23}), (iii) Magnetic field strength ($B_s$)= 3000~G, and (iv) Polarity separation between two poles of the BMR is three times the radius of a pole.
No nonlinearity is imposed in this case.

\subsubsection{STABLE as dynamo model} 
\label{sec:dyn}
By default, STABLE was designed to be used as a dynamo model in which the SpotMaker algorithm plays a crucial role.
This algorithm deposits the Bipolar Magnetic Regions (BMRs) on the surface based on the toroidal field at the base of CZ. 
When this field exceeds a (assigned) threshold value and the time delay between the two successive spots is greater than $\Delta$, then this algorithm adds a spot on the surface.  The delay $\Delta$ is taken from an observed distribution of the time delay of BMRs; see Sec. 2 of \citet{KM17}.  
Once the timing of the spot is decided, the flux and tilt of the spots are taken from their observed distributions (Equations 8, 9, and 10 of \citet{KM17}). 
The distribution of fields in the spot and the internal profile are the same as discussed in the previous section.
We note that in this case, to limit the growth of the magnetic field in dynamo, a magnetic field-dependent quenching in the tilt angle (of the form $1 / (1 + (B / B_{\rm sat})^2)$, where $B$ is the average $B_\phi$ at BCZ and $B_{\rm sat} = 100$~kG) is included as inspired by observations \citep{Jha20, sreedevi24}.

Once the model deposits BMRs on the solar surface, the decay and dispersals of the field produce a poloidal field. This poloidal field, through differential rotation, produces a toroidal field, which again gives rise to BMR eruptions and continues the dynamo loop; see, e.g., \citet{KM17, KM18} for more details of the model and the representative dynamo solutions. 

\section{Results and Discussion} 
\label{sec:results}
Our work analyzes the behavior of magnetic field evolution by varying the meridional flow speed $v_0$.
First, we compare the poloidal field evolution from the 3D dynamo model (STABLE) with that of the 2D model incorporating the \bl\  process by an $\alpha$ term and check which model is consistent with the SFT model. 
\begin{figure}
     \includegraphics[width=\columnwidth]{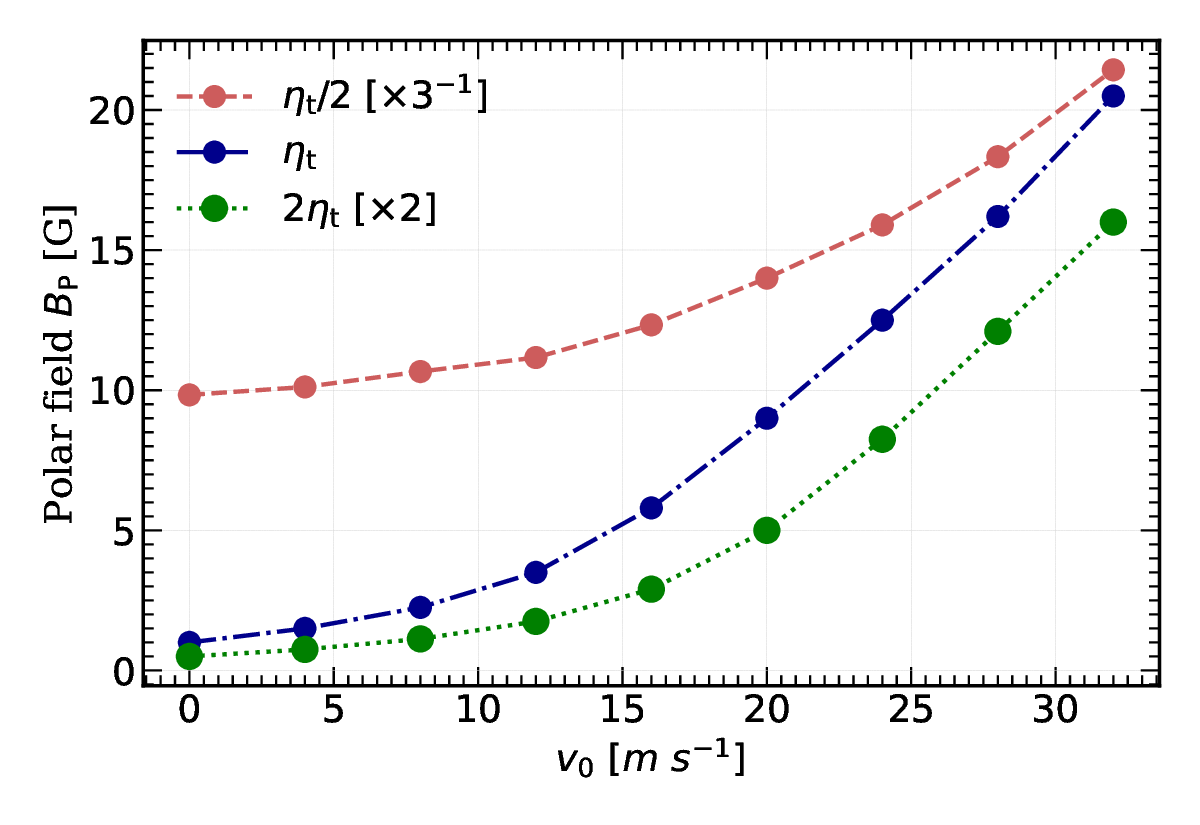}
    \caption{Dependence of saturated 
    radial field on meridional circulation amplitude $v_0$ in the 2D dynamo model with explicit $\alpha$ term for \bl\ process at different values of diffusivity; blue: the reference diffusivity $\eta_{\rm t}$ as given by \Eq{eq:eta}, green: $2\eta_{\rm t}$ (scaled by a factor of 2 to fit in the range), red: $\eta_{\rm t}/2$ (scaled by a factor of 1/3).}
    \label{fig:2d_pf}
\end{figure}

\subsection{Polar field evolution in 2D model with $\alpha$ term}
To examine the results from a 2D model incorporating the \bl\ $\alpha$ effect, we perform several simulations at different values of \mf\ speed $v_0$, starting from 0 to 32 \mps\ and compute the average value of the surface radial field from  $75^\circ$ latitude to the north pole. We define this polar field as $B_{\rm P}$.
This radial field first increases and then slowly decreases due to the supply of more leading (opposite) polarity flux towards the pole. Finally, it tends to saturate, as seen in Figure~10 of the Appendix. We note that if we do not include the radial magnetic pumping, then the diffusion of the poloidal field across the surface is high, and the $B_{\rm P}$ cannot saturate in the later stage; see \citet{KC16} Fig 5.
In the present case, when the polar field (approximately) saturates in time, we take its value and plot it for different \mc\ speeds as shown in \Fig{fig:2d_pf}. 
In this model, we find that the polar field consistently increases with meridional flow speed. 
This increasing trend is due to the fact that 
in this model, there is a large separation between the leading and trailing polarity 
fluxes. And, since the meridional circulation is weak near the equator, increasing the flow predominantly enhances the transport of the trailing polarity flux towards the pole (see Figure 10 in the Appendix for a pictorial representation of the radial field).
This causes a monotonous increase of the polar field with $v_0$, which is strikingly in contrast to the expectation of the SFT model.
Even by changing the latitudinal extent of the $\alpha$ term, or by placing $\alpha$ at different latitudes, and varying the diffusivity (see dash and dotted lines in \Fig{fig:2d_pf}),  we observe a similar increasing trend in the polar field with the flow.

\subsection{Polar field evolution in STABLE  with single BMR}

\begin{figure}
    \centering

    \includegraphics[width=0.98\columnwidth]{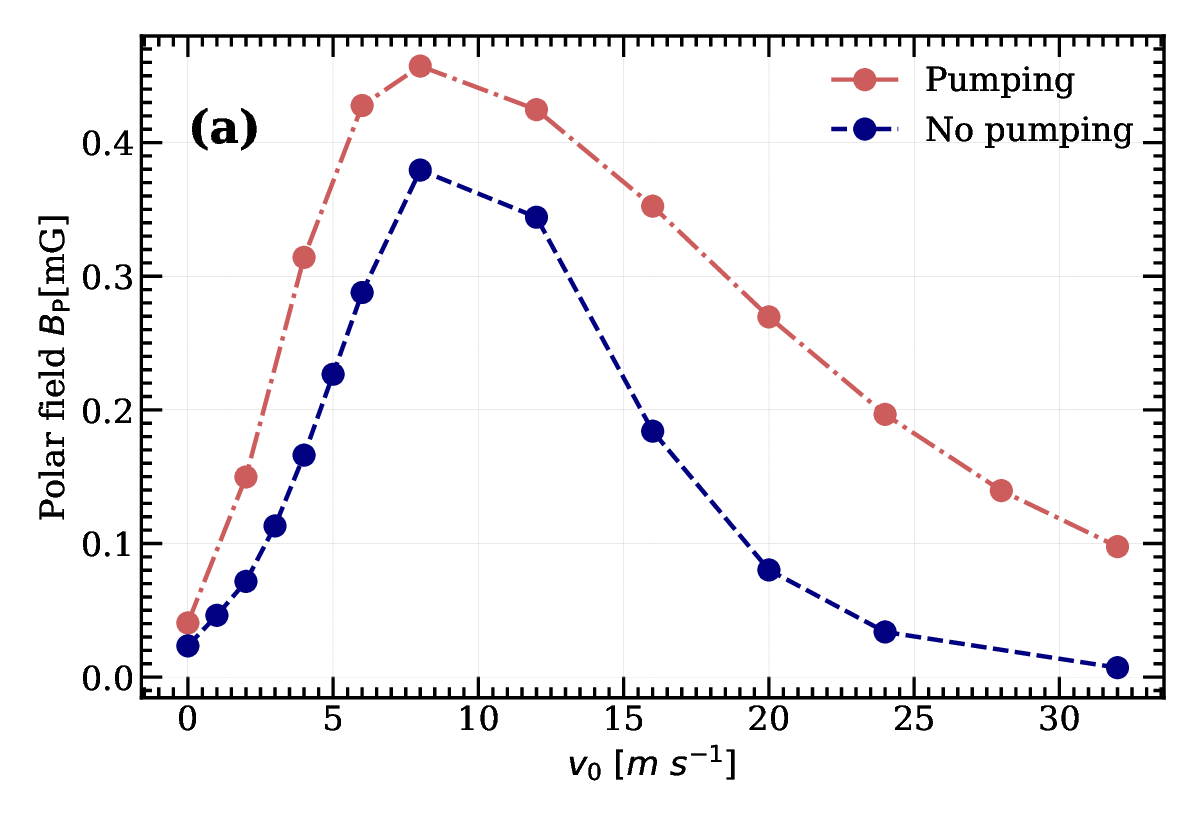}
    \caption{Variation of saturated polar field strength $B_{\rm P}$
    %(b) dipole moment $D$ 
    with the flow amplitude $v_0$, from a simulation in which a single BMR is placed at $15^{\circ}$ latitude. Red and blue colors correspond to simulations with and without the magnetic pumping. The data for latter case are multiplied by a factor of 60 for better visibility.}
    \label{fig:lat15_pf_df_p}
\end{figure}

Now, we observe the behavior of the polar field in the STABLE model in which a BMR is included at $15^\circ$ latitude in the northern hemisphere, and no $\alpha$ term is added in the equation (\Sec{sec:STABsingle}).  The decay of this BMR generates a poloidal field.  We again perform the simulations at different values of \mf\ speed $v_0$ starting from 0 to 32 \mps\ and compute the average value of the polar field from $75^\circ$ latitude to the north pole.
Contrary to the results obtained from the 2D model with explicit $\alpha$ prescription, the STABLE model reveals a distinct trend.
As seen in \Fig{fig:lat15_pf_df_p}, the polar field initially increases, and then beyond a certain value of $v_0$, it decreases. 
The polar magnetic field, $B_{\rm P}$, is extremely weak at $v_0=0$ because, in this case, the leading and trailing polarity fluxes largely cancel each other, and there will be a negligible trailing polarity field that goes to the pole. With the increase of $v_0$, the polar field initially increases due to the 
transport of more trailing polarity flux\footnote{Both leading and trailing polarities move towards the pole, but as the trailing one appears at the higher latitude, it experiences more drag of the flow.} 
%enhanced transport of trailing polarity flux 
towards the pole, facilitating the cross-equatorial cancellation of the leading polarity.

However, as $v_0$ continues to boost up further, the meridional flow also efficiently transports the leading polarity to the pole, and thus, the cancellation of the leading polarity flux across the equator reduces, resulting in a decrease in $B_{\rm P}$. 
The most intense polar field is generated at a moderate meridional flow (at $v_0 = 8$ \mps\ in this configuration). 
Contrary to the 2D dynamo model that incorporates the \bl\ $\alpha$ effect, the STABLE model exhibits a smaller separation between trailing and leading polarities, resulting in both polarities moving towards the pole; see Figure 11 in the Appendix for a pictorial representation of how a BMR evolves in our 3D STABLE model. 
Interestingly, in the 2D model with $\alpha$ parameterization for \bl\ process, it is primarily the trailing polarity flux that moves toward the pole; see Figure 10 in the Appendix.

The observed trend of the polar field with the meridional flow speed in STABLE is consistent with the expectation from SFT models, especially with the works of \citet{Bau04, Schri08}. By incorporating double rings as the replacement of the \bl\ $\alpha$ effect in a dynamo model, \citet{munoz10} produced a decreasing trend of the polar field with the meridional flow. However, they did not show the behavior at low meridional flow speed.  Thus, the increase of the polar field at low meridional circulation and the turning point were unexplored. 

Our study marks the first instance of a dynamo model demonstrating such a variation between the polar field strength and meridional flow as found in SFT models. Moreover, \citet{Pawan2024} showed that the STABLE model successfully reproduces the variation of the eventual dipole moment with the latitude of BMR as found in SFT models.  These suggest that the results generated by the STABLE dynamo model are reliable.
%, as they successfully replicate the results observed in SFT models. 

We recall that in our 3D dynamo model, we have used downward magnetic pumping, which suppresses the diffusion of the magnetic field through the surface \citep{Ca12, KC16}. Thus, when we remove the pumping, we observe that the strength of the polar field is reduced. However, the overall profile of the polar field vs meridional flow remains almost unchanged; see \Fig{fig:lat15_pf_df_p}. 

Thus, the variation of the polar field with the meridional flow, as found in our STABLE model, is robust. Therefore, we shall present the results of the polar field by including magnetic pumping only.

\begin{figure}
    \centering
    \includegraphics[width=0.98\columnwidth]{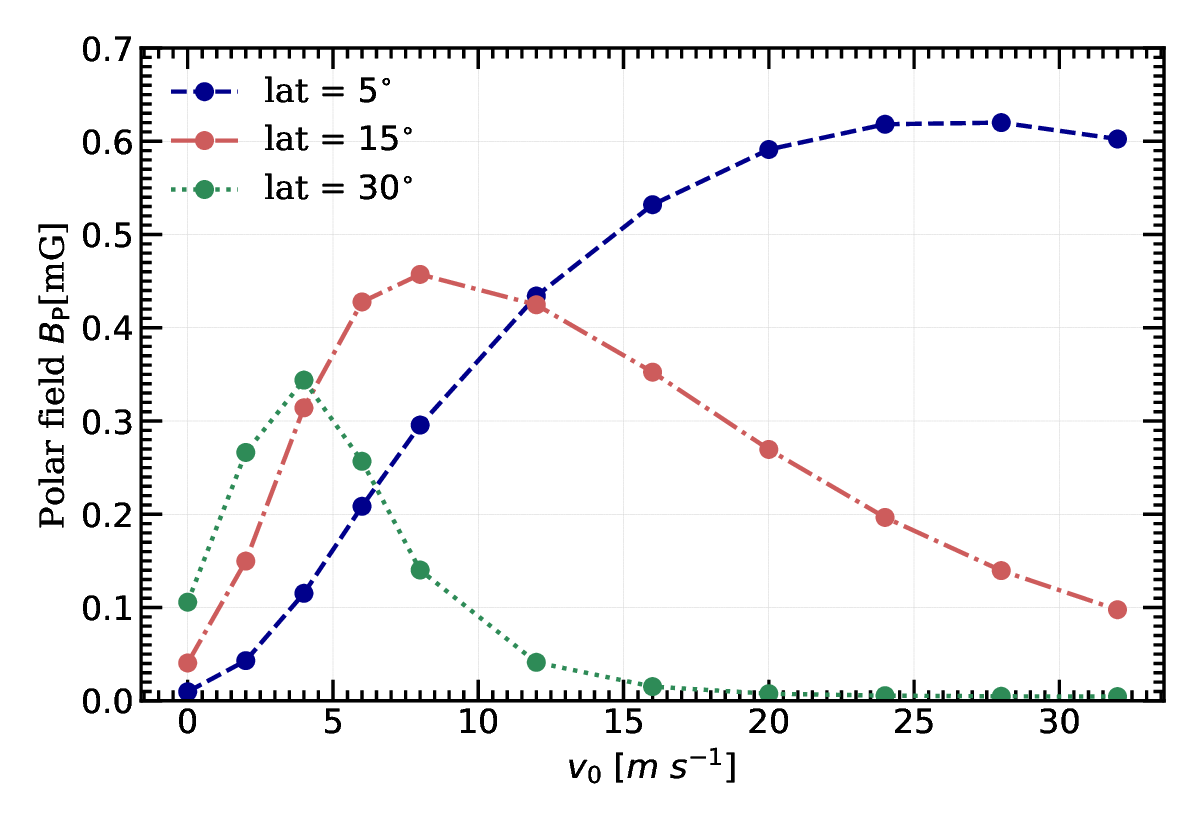}
    \caption{Variation of the polar field strength $B_{\rm P}$  with the flow amplitude $v_0$ for three different cases in which the BMR is placed at $5^{\circ}$ (blue), $15^{\circ}$ (red), and $30^{\circ}$ (green) latitudes.
    }
    \label{fig:pf_d_alllat}
\end{figure}

In Sun, the BMRs are produced at different latitudes while the meridional flow peaks around the mid-latitude. Hence, the BMRs appearing near the equator will experience weak flow as compared to mid-latitude. The same BMR, when placed at different latitudes, will produce different amounts of polar field. To demonstrate this, in \Fig{fig:pf_d_alllat}, we show the variation of the polar field with the meridional flow of the same BMR placed at $5^{\circ}, 15^{\circ}$ and $30^{\circ}$ latitudes.
From this figure, we clearly observe a shift in the peak of the polar field towards higher ranges of magnetic field strength and $v_0$ with the decrease of BMR latitude.

This shift towards a higher polar field occurs because a BMR placed closer to the equator experiences more cross-equatorial cancellation, resulting in efficient transport of trailing polarity flux towards pole. 
This feature is supported by the findings of \cite{KM18, Kar20} and also consistent with several previous SFT models \citep{JCS14, Pet20}. 
The shift of the peak of the polar field towards higher meridional flow in \Fig{fig:pf_d_alllat} also occurs because 
the meridional flow speed is weak at low latitude 
and the flow needs to be increased enough to cause the decrease in the polar field strength when a BMR is placed at low latitude (compare with \Fig{fig:lat15_pf_df_p}). 

\begin{figure}
    \centering
    \includegraphics[width=\columnwidth]{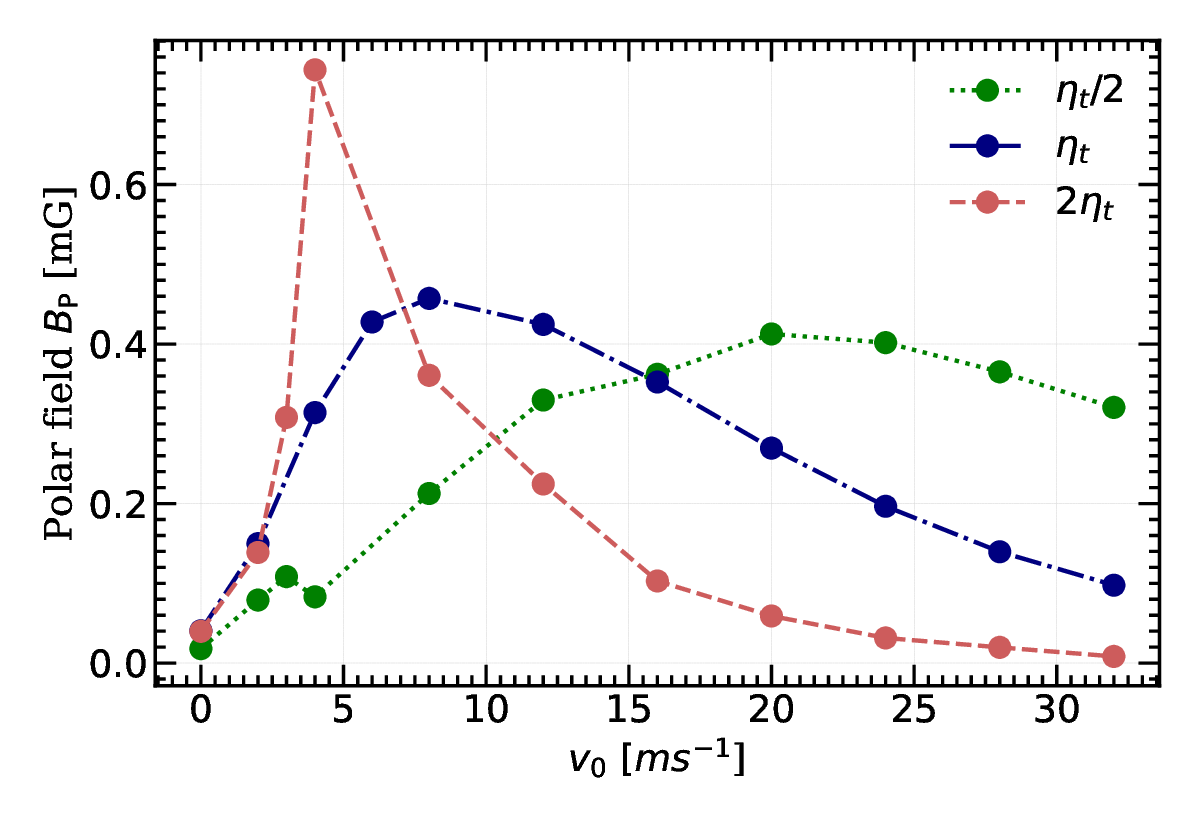}
    \caption{The polar field strength $B_{\rm P}$
    vs the meridional flow amplitude at three distinct diffusivity values; blue: $\eta_t$---the reference diffusivity as given by \Eq{eq:eta}, red: $2\eta_t$, green: $\eta_t/2$. Here, a single BMR is placed at $15^{\circ}$ latitude.}
    \label{fig:diff}
\end{figure}

To further investigate the sensitivity of the polar field variation to the model parameters, we perform two more sets of simulations, one at double and another at half of the reference diffusivity as given by \Eq{eq:eta}. The reference case is the one where a single BMR is placed at $15^{\circ}$ latitude.
As illustrated in \Fig{fig:diff}, in both 
cases, the polar field vs meridional flow exhibits a similar trend. However, with the increase of diffusivity, the peak shifts towards the lower $v_0$. In the case of higher diffusivity (red curve), the cross-equatorial diffusion is high, and even a slight increase in meridional flow can swiftly transport the trailing polarity towards the polar region, resulting in an increased strength of the polar field. Further increase of flow leads to enhanced transport of both polarities, which causes less net polar flux. Moreover, in the case of high meridional flow, increased diffusivity results in a decrease in polar field strength.
Interestingly, 
%in contrast to the 2D model, 
in this scenario, high diffusivity enhances the efficiency of polar field generation, thereby increasing the peak of the polar field, which contradicts the observations from the 2D model (\Fig{fig:2d_pf}).

\subsection{Toroidal field evolution in STABLE  with single BMR}
\begin{figure}
    \includegraphics[width=\columnwidth]{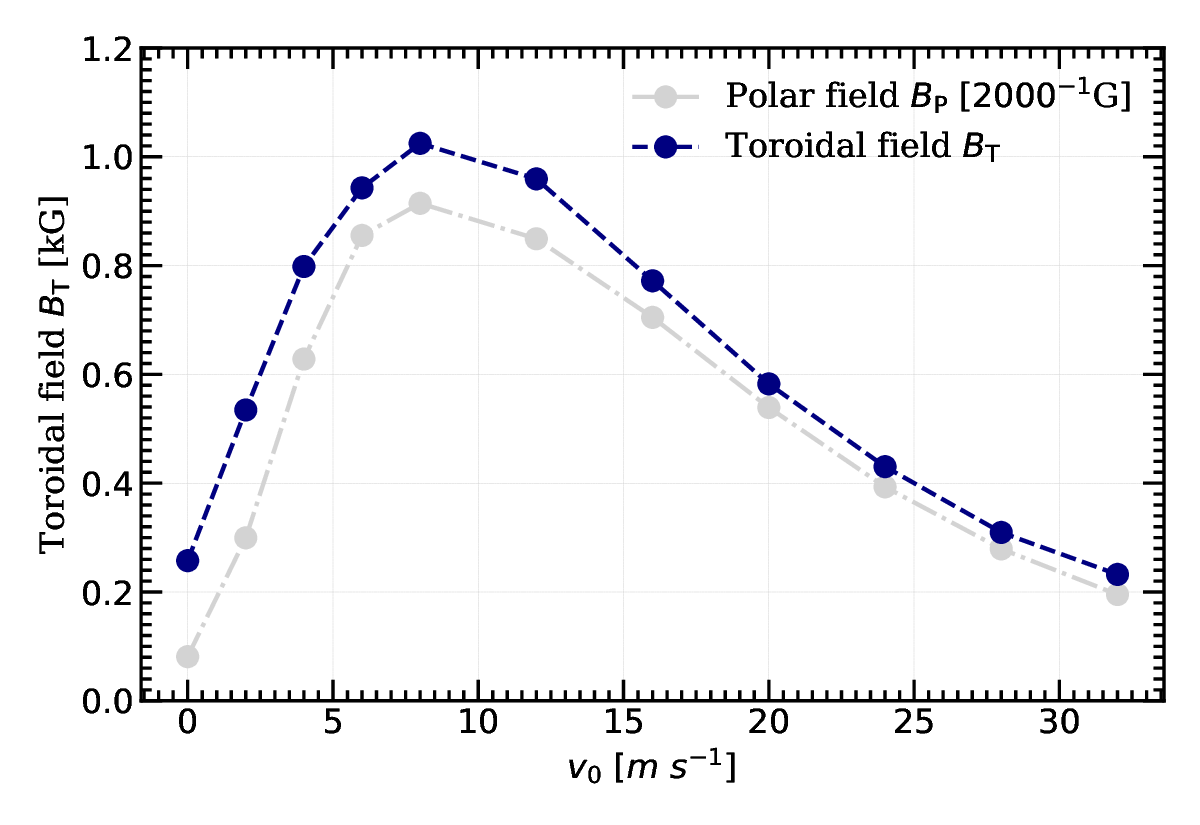}
    \includegraphics[width=\columnwidth]{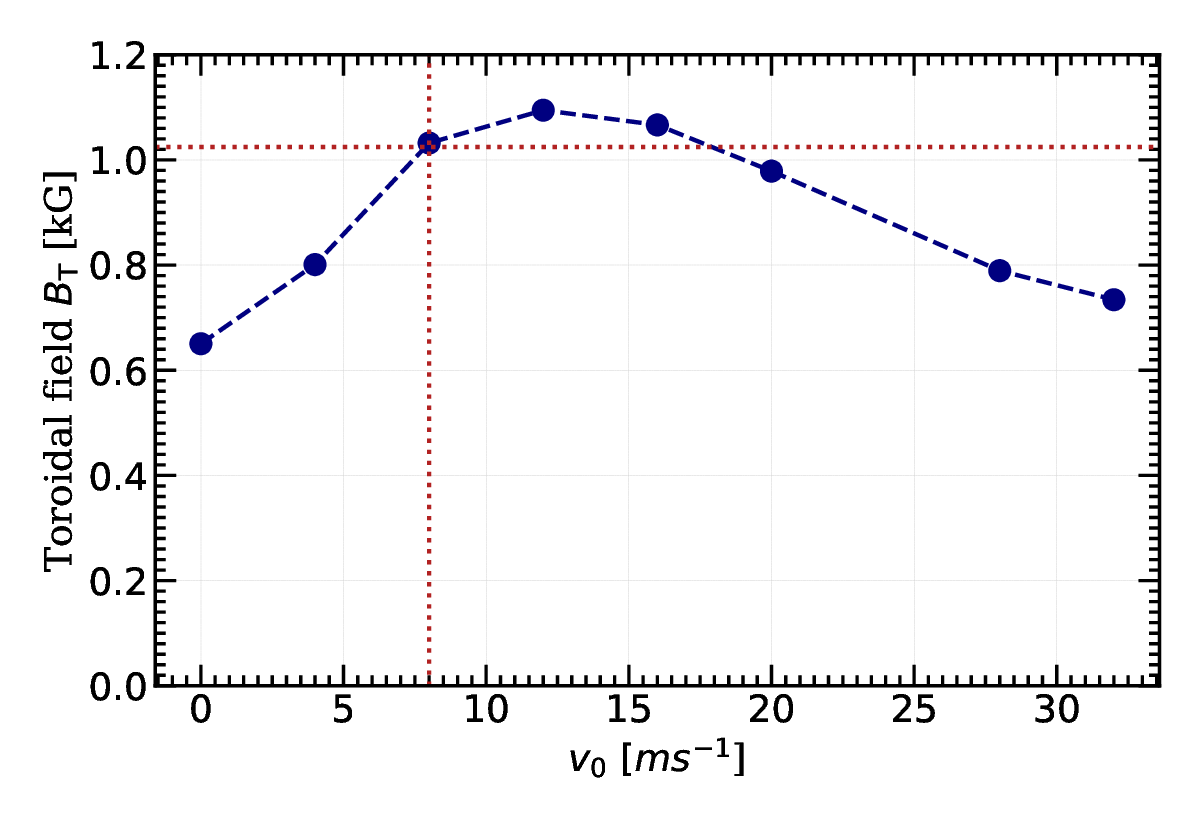}
    \caption{
    Top: Representation of the toroidal field $B_{\rm T}$ (blue curve) in comparison with the poloidal field $B_{\rm P}$ (gray curve, scaled by a factor of 2000) in relation with the flow amplitude.
    Bottom: Same as the top but the toroidal field variation from a set of simulations in which the same initial magnetic field is taken from the output of the simulation $v_0 = 8$ \mps. }
    \label{fig:omeff}
\end{figure}

\begin{figure*}
    \centering
    \includegraphics[width=\columnwidth]{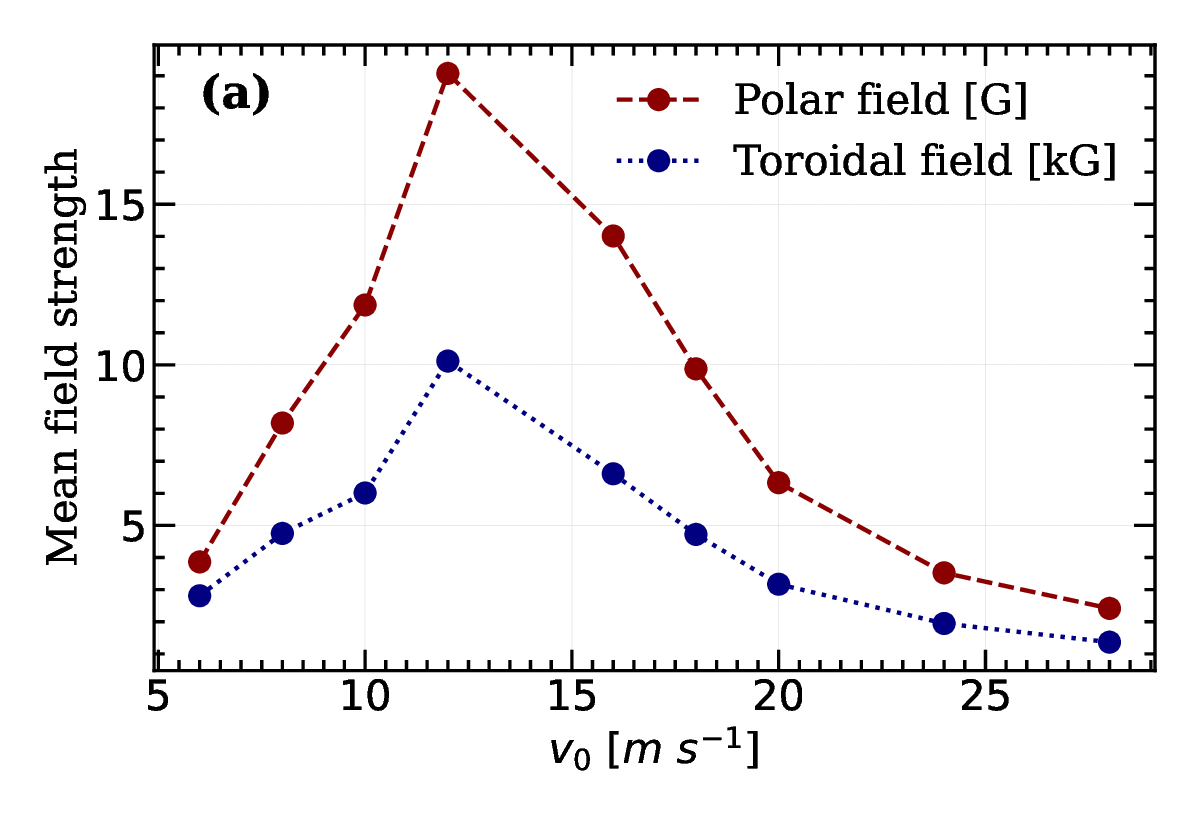}
    \includegraphics[width=\columnwidth]{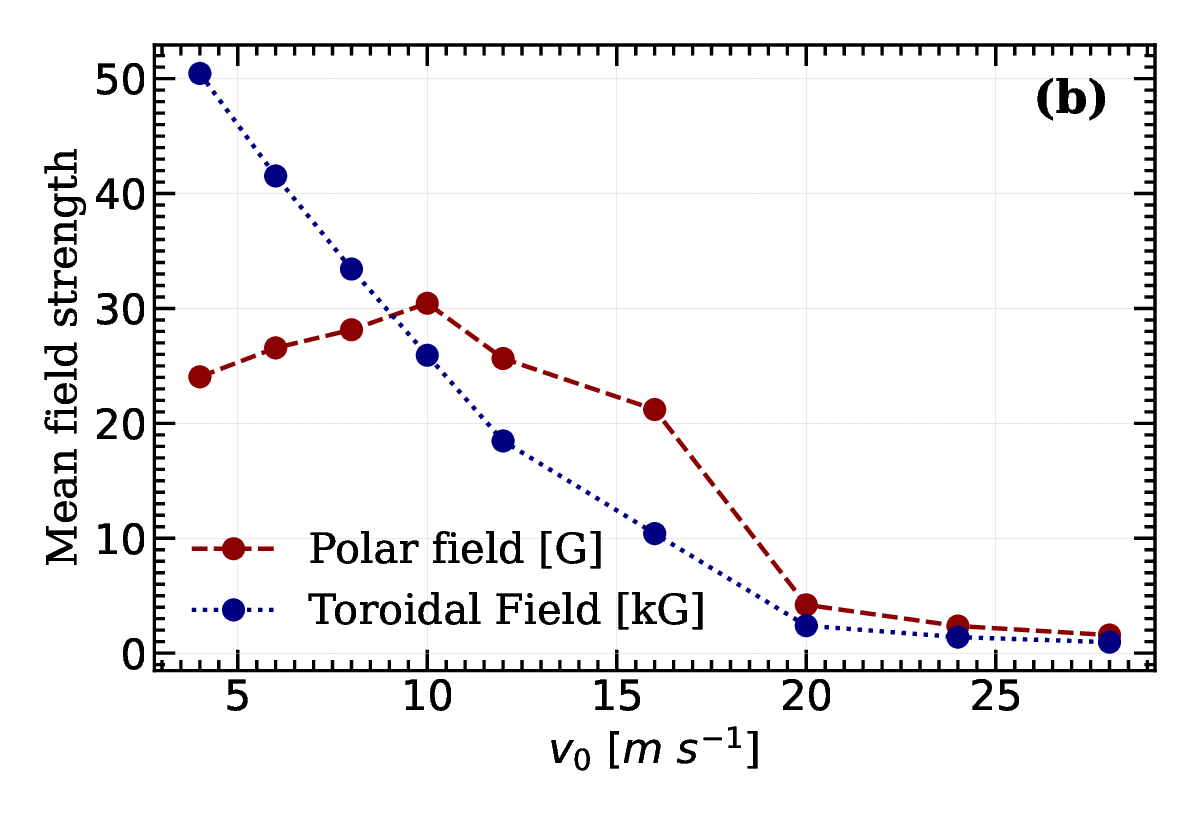}
    \includegraphics[width=\columnwidth]{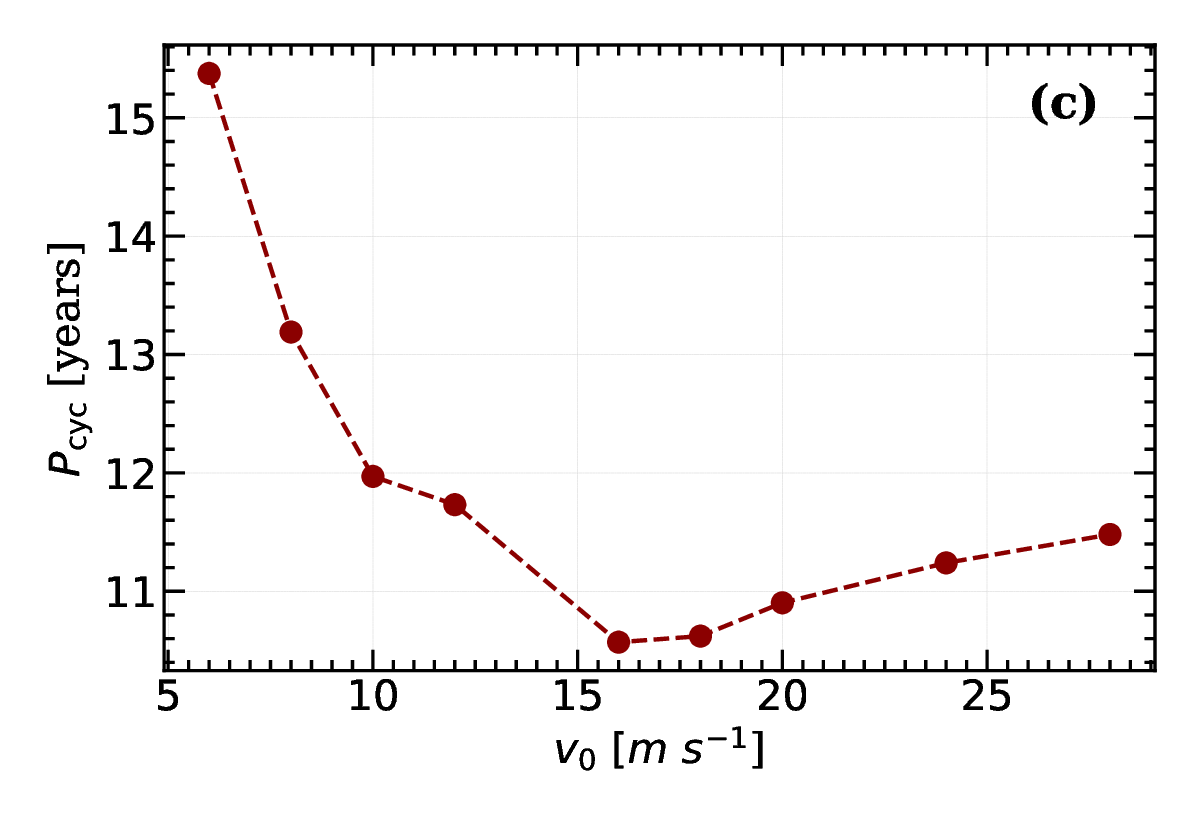}
    \includegraphics[width=\columnwidth]{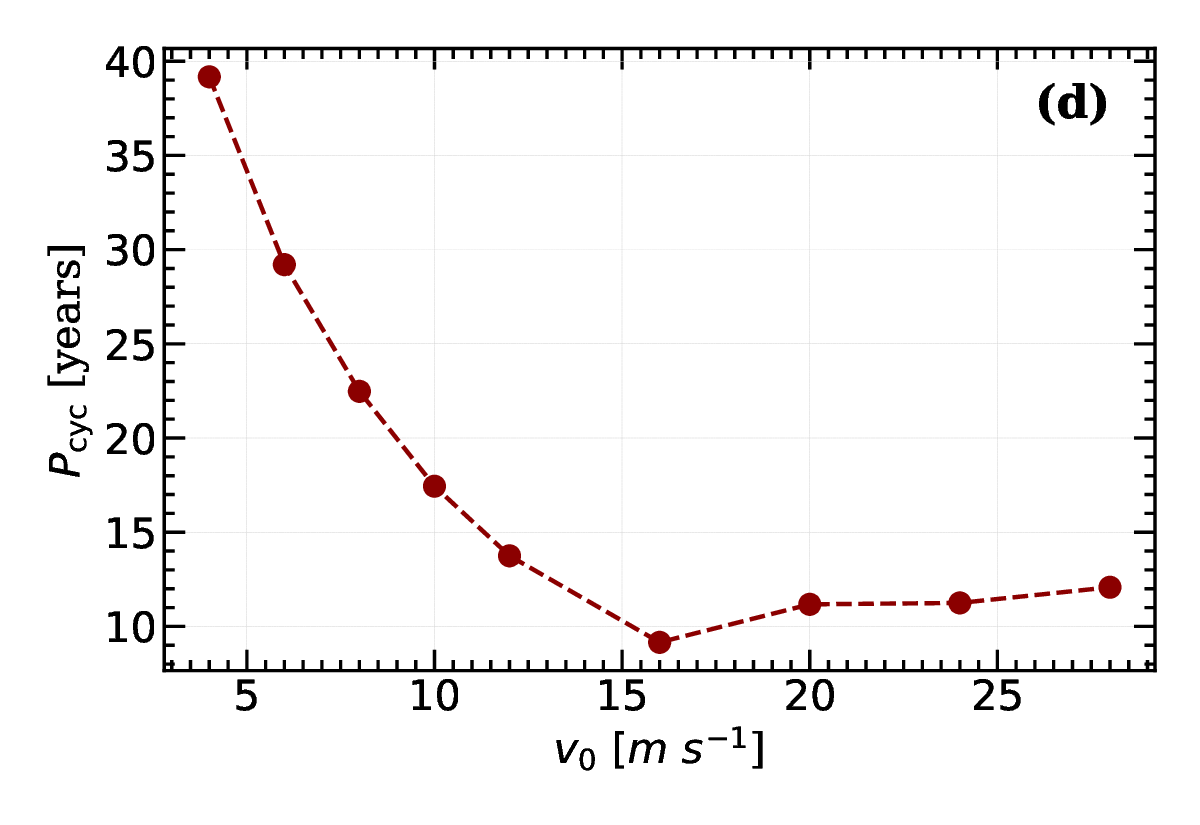}
    \caption{
    Variation of the (a) mean magnetic fields and (c) cycle period with the flow speed. The right panels are same as the left ones but from a low diffusivity case ($\eta_t/2$).
    }
    \label{fig:dyn}
\end{figure*}

While the polar field is measured on the solar surface and thus can be compared with observations \citep[e.g,][]{Mord22}, the toroidal field is the one that gives the sunspots and thus determines the strength of the solar cycle. Therefore we now examine the variation of the toroidal field at 
different speeds of meridional flow.
To do so, we first calculate the average value of the axisymmetric toroidal field $\langle B_\phi \rangle $ (azimuthal average) from $30^\circ$ latitude to the equator. We denote this quantity as $B_{\rm T}$, which slowly increases in time and then tends to saturate. The value of $B_{\rm T}$ is taken at the same time when the saturated polar field is computed. 
We first observe the toroidal field from the set of simulations in which we had deposited the BMR at $15^\circ$ north, and is shown in \Fig{fig:omeff}, top panel. 

We observe a very similar trend in the toroidal field as found for the polar field (shaded line). 
One might expect the toroidal field, which is generated from the poloidal one, to follow the same trend as the poloidal field. 
However, it is not that obvious.
The generation of the toroidal field is influenced  
by an additional physics as identified by \citet{YNM08} and later demonstrated in different context by \citet{Kar10} and \citet{KC11}. 
If the meridional flow is increased, then the cycle period becomes short, and this causes two competing effects: (I) the shear gets less time to induce the toroidal field, thereby reducing toroidal field strength, and (II) the magnetic fields remain at the base of CZ for a shorter time, and thus the diffusion gets less time to diffuse the fields. As a result, toroidal field strength increases.
In the high diffusivity regime (as in our case), effect II dominates as long as meridional flow is moderate. However, when meridional flow becomes too high, the shearing effect (effect I) becomes dominant, and the toroidal field tends to decrease with the increase in flow speed.   
Therefore, for the same initial poloidal field, the toroidal field strength should first increase at low meridional flow speed, reach a maximum, and then decrease with the increase of meridional flow speed. However, in \Fig{fig:omeff} (top panel), we see almost a similar trend in the toroidal field as seen for poloidal one (gray points) because the poloidal field is changing in each simulation of different meridional flow speeds.

To show the effect of meridional flow alone on the toroidal field, we take the saturated poloidal field from the run of $v_0 = 8$ \mps\  and perform a new set of simulations with this same poloidal field as an input. We note that in this new set of simulations, no BMR is deposited, and thus, the initial poloidal field (which is concentrated near the pole) remains the same.  
In  \Fig{fig:omeff}, bottom panel, we observe that the toroidal field produced from this set of simulations shows the variation as expected from the above explanation (competing effects of shear and diffusion). We clearly see that the toroidal field is increased by some amount from the value at $v_0 = 8$ \mps\ (shown by dotted line) up to about $v_0 = 15$ \mps\, and then it decreases.
However, this effect of the meridional flow on the toroidal field is less than the effect on the generation of the poloidal field, and that is why we do not observe two distinct peaks in \Fig{fig:omeff}.

The above two sets of simulations demonstrate that the meridional flow can determine the strength of the solar cycle by regulating the generation of the poloidal field on the surface (\bl\ process) and by affecting the diffusion of the toroidal field in CZ. This motivates us to explore the strength of the magnetic cycle in dynamo simulations at varying meridional velocity. 

\subsection{Magnetic cycles in STABLE dynamo model}

We now perform a set of dynamo simulations at different values of \mf\ speed using the STABLE in default (dynamo) mode. In this set of simulations, we include a weak dipolar field for the initial condition and run the model for several cycles until it produces stable magnetic cycles. We note that in these STABLE dynamo simulations, BMRs are spontaneously produced based on the toroidal field at BCZ and the decay of which generates a poloidal field. We further note that the magnetic field here is stabilized because of the magnetic field-dependent quenching, as mentioned in \Sec{sec:dyn}. The strength of the saturated magnetic cycle as measured by the average value of the toroidal field at $r = 0.72R_{\rm s}$ and $\theta$ from $30^{\circ}$ to the equator from different simulations at varying \mc\ is shown in \Fig{fig:dyn}(a).
We observe that the cycle strength initially increases, reaches a maximum, and then declines with the increase of $v_0$. This behavior is attributed to the combined effect of meridional flow on the generation of a poloidal field and the diffusion of the magnetic field, as explained in the above section. 
%This non-monotonous behavior of magnetic cycle with \mc\ is in striking contrast to the high diffusivity flux transport dynamo models with \bl\ $\alpha$ process \citep{Kar10, munoz10}. 
However, when the diffusivity is reduced by half, the diffusion effect will be less dominant over the shearing effect, and thus, increasing $v_0$ gives less time for the shear to generate a toroidal field. This causes a consistent decrease of the toroidal field with $v_0$ as seen in \Fig{fig:dyn}(b).

Another interesting trend is seen in the cycle period for which we expect a monotonically decreasing trend with the increase of flow speed in traditional flux transport dynamo model with \bl\ $\alpha$ \citep[see Fig. 4 of][]{YNM08,Kar10, munoz10}. Here, in contrast, we observe a slight increase in the cycle period at high meridional flow; see \Fig{fig:dyn}(c). This trend remains more or less consistent even at lower diffusivity value; see \Fig{fig:dyn}(d).
The slight increase in cycle period at high $v_0$ (beyond about 16 \mps) is because the generation of the poloidal field weakens at fast meridional flow  (see \Fig{fig:dyn}(a)) and the poloidal field needs more time to reverse the old field.
\begin{figure}
    \centering
    \includegraphics[width=\columnwidth]{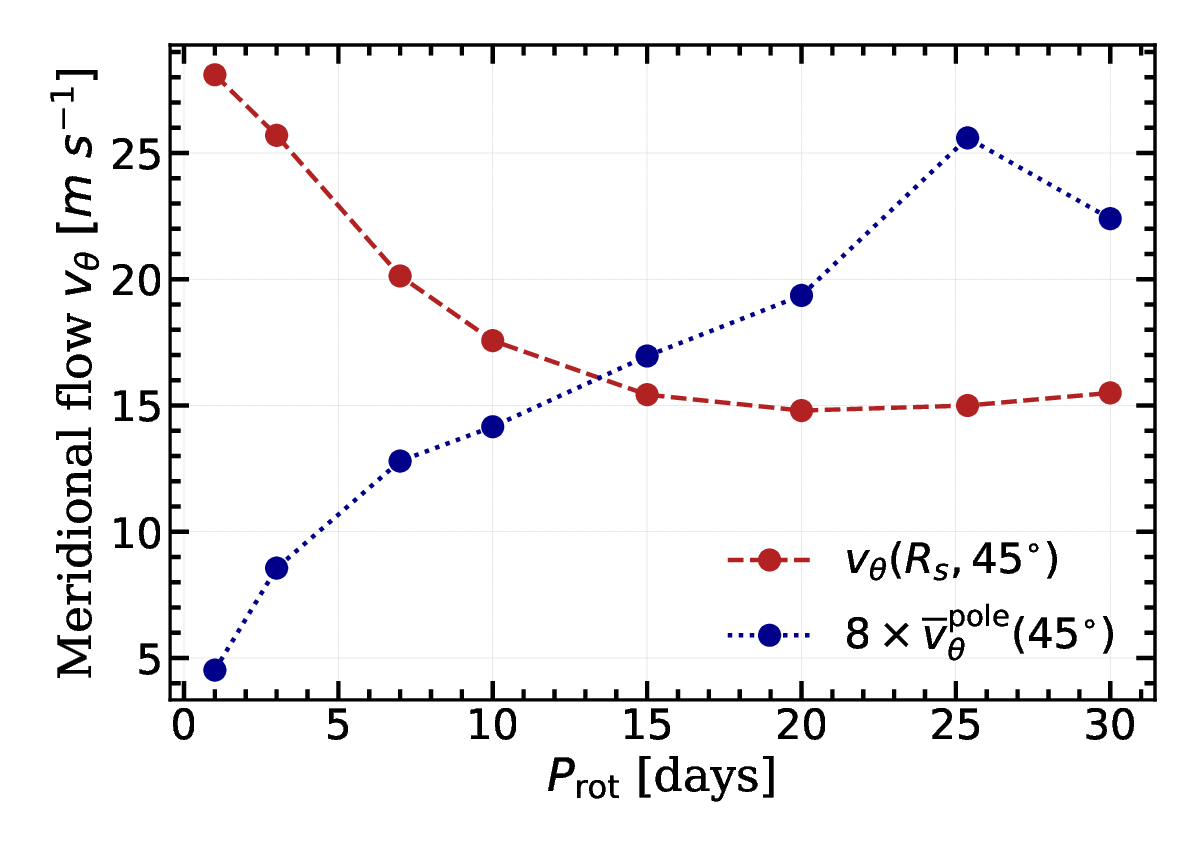}
    \caption{Variation of  
    %the maximum $v_{\theta}$ 
    %(red dashed curve) at 45$^\circ$ latitude 
    the latitudinal component of the surface meridional flow $v_{\theta} (R_s, 45^\circ)$ 
    (red dashed curve) and the {\it radial} average of the poleward flow $\overline{v}_{\theta}^{\rm pole} $ at $45^\circ$ latitude (blue dotted curve, scaled by a factor of 8) with the rotation period of the stars.}
    \label{fig:vtheta}
\end{figure}
\subsection{Implication to stellar cycles}
The above sets of simulations teach us that the magnetic cycle amplitude and duration are largely determined by the meridional flow speed. Features of the stellar cycle vary with the rotation rate of the star \citep{Baliu95, garg19}. Particularly, the cycle strength increases with the increase of rotation rate and then saturates at rapidly rotating stars \citep{Noyes84a, wright11}. While the cycle period shows a bit complicated trend at rapidly rotating stars, an increasing trend with the rotation period is identified for slow rotators \citep{BoroSaikia18}. 
We also note that, with the rotation rate, the meridional circulation strength changes in the solar-like stars. 
Mean-field hydrodynamic models and convection simulations show that the rapidly rotating stars are characterized by a weak meridional flow, which is confined only near the boundaries, while the slowly rotating stars possess a strong flow in the bulk of the CZ; see Fig. 5 in \citet{KO12b}. 
Therefore, the question is how much is the role of \mf\ in determining the stellar cycle amplitude and duration. 

To address the above question, we conduct a series of stellar dynamo simulations by taking the meridional flow from \citet{KO12b} model for $1 M_{\rm sun}$ stars of rotation periods of 1, 3, 7, 10, 15, 20, 25.38 (solar value), and 30 days. 
The radius of these stars are 0.896, 0.898, 0.902, 0.907, 0.920, 0.949, 1, 1.08 $R_{\rm s}$.
The variations of the latitudinal component of the surface meridional circulation $v_\theta (R_s, 45^\circ)$ and the {\it radial} average of the poleward flow at mid-latitude are illustrated in \Fig{fig:vtheta}. We observe that the rapid rotators exhibit higher velocities near the surface but lower flow in the bulk of the CZ. Conversely, the slow rotators posses a strong flow within the bulk but are weak on the surface. In other words, the flow amplitude near the surface steadily decreases with the rotation period, whereas the flow within the bulk of the convection zone tends to increase with rotation period.

\begin{figure}
    \centering
    \includegraphics[width=\columnwidth]{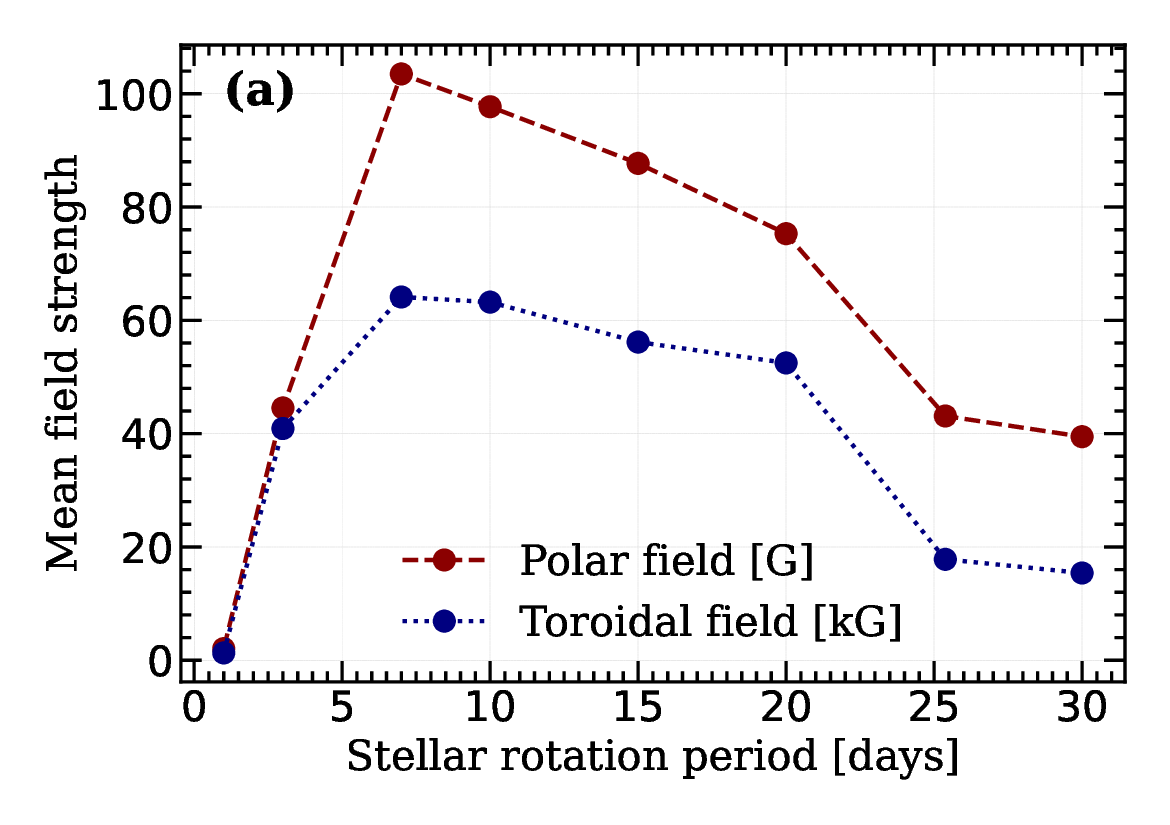}
    \includegraphics[width=\columnwidth]{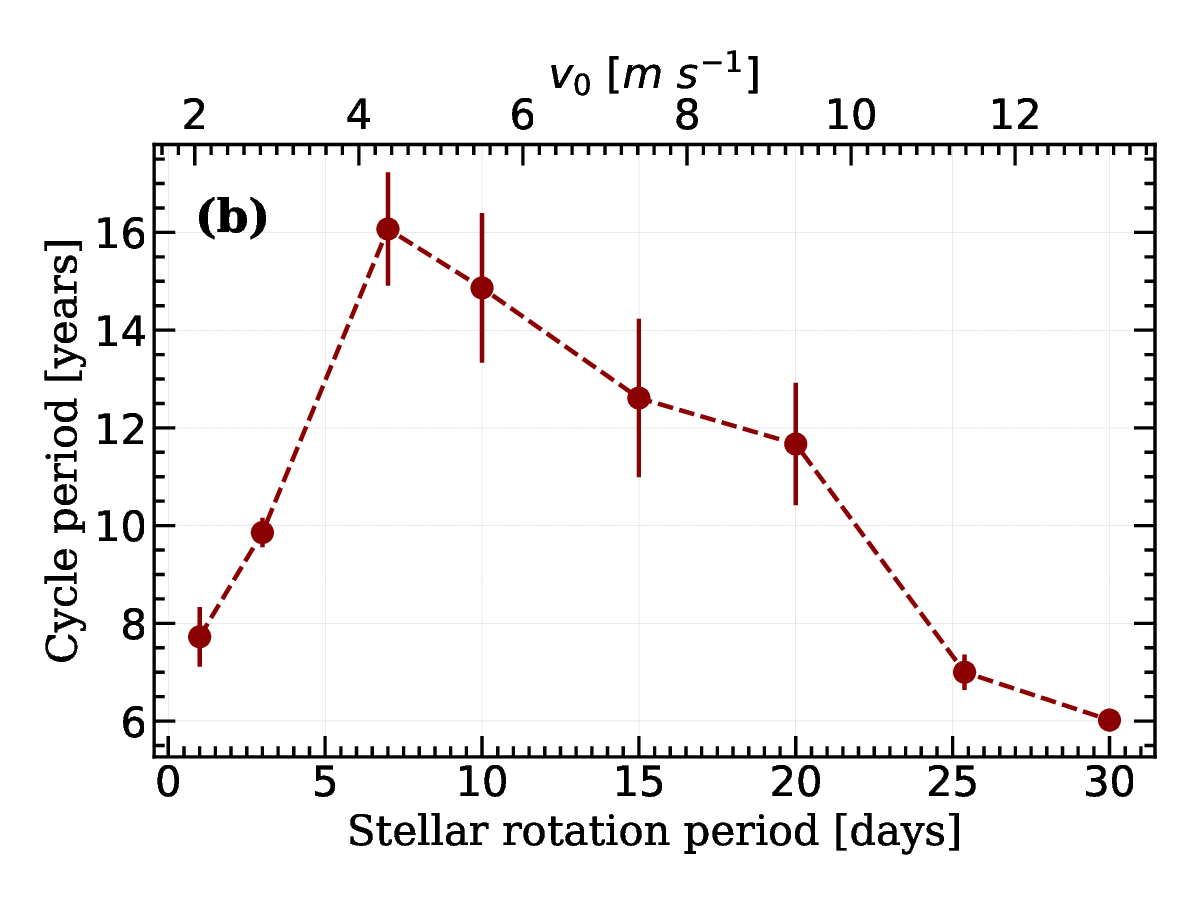}
    \caption{Same as \Fig{fig:dyn}, but from the set of simulations in which the meridional flow is taken from \citet{KO12b} mean-field hydrodynamics model of sun-like stars having rotation periods varying from 1--30 days. The horizontal axis represents the rotation period of the corresponding mean-field model of stars.}
    \label{fig:stel_dyn}
\end{figure}

\begin{figure}
    \centering
    \includegraphics[width=1.1\columnwidth]{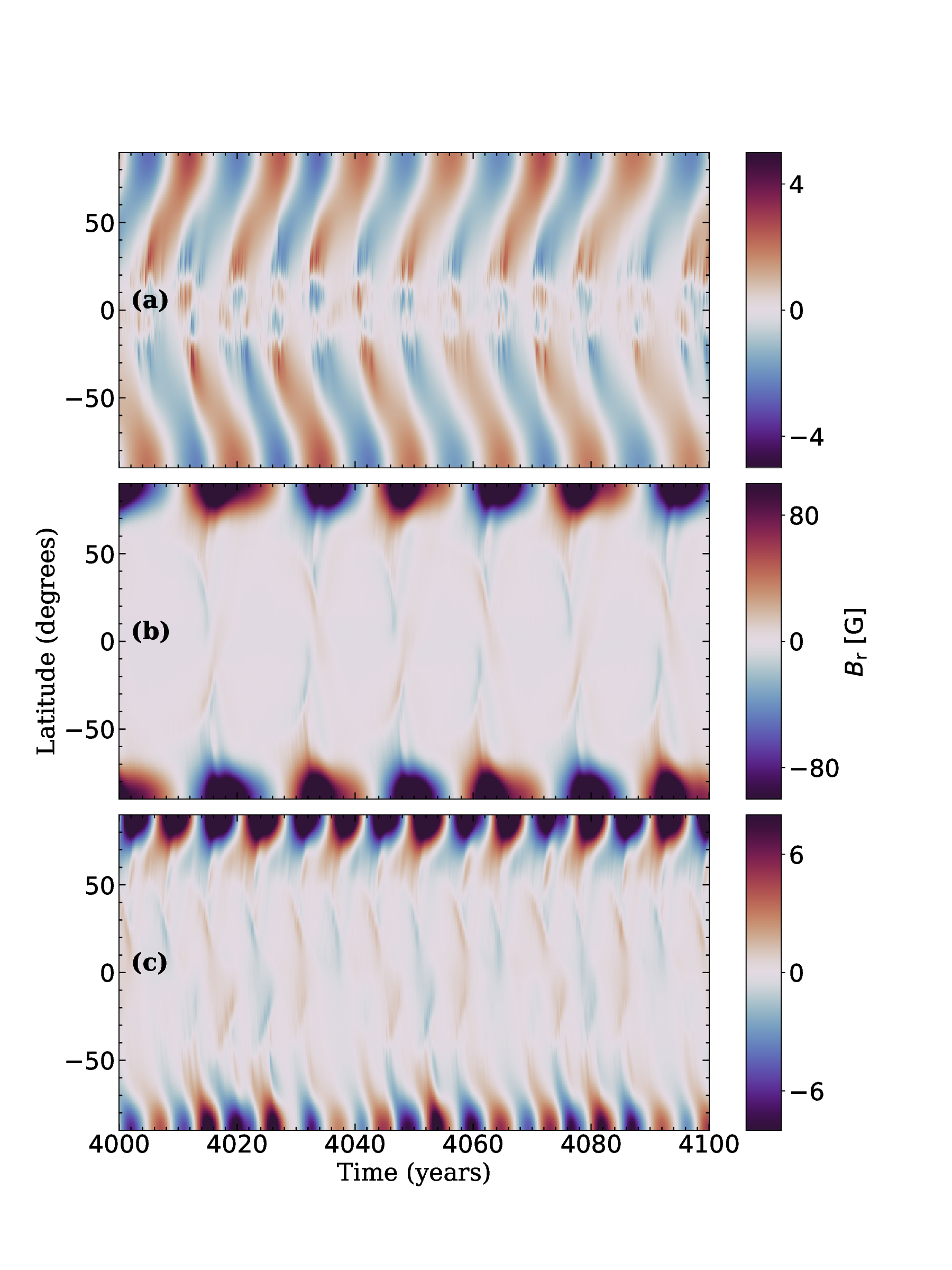}
    \caption{Time-latitude distribution of the surface radial 
    magnetic field $B_{\rm r}$ [in G] for stars of (a) 1 day, (b) 7 days, and (c) 25.38 days (solar value) rotation periods.}
    \label{fig:stel_bfly}
\end{figure}
%The time-latitude plot, as shown in  \Fig{fig:stel_bfly}, also clearly shows the effect of meridional flow in different ranges of stars.
Now we analyse the poloidal and toroidal fields obtained from these simulations which are shown in \Fig{fig:stel_dyn}(a). We see that both the fields initially increase, reach maximum, and then decline with the increase in the rotation period. 
For a rotation period of 1 day, the surface meridional flow is very strong, while the flow within the bulk of CZ is negligible; both of these contribute to an extremely weak polar field. As the rotation period increases, the meridional flow within the bulk (as shown in \Fig{fig:vtheta}) strengthens, resulting in an increase in the polar field. On the other hand, at a rotation period of 30 days, the bulk flow is strongest, but the surface flow is weakest, leading to a weak polar field again.
When the rotation period decreases from this point, the surface flow intensifies, contributing to a stronger polar field. Therefore, an increase in the polar field is observed when transitioning from low to high meridional flow in bulk (from fast to slow rotators) or from low to high surface flow (from slow to fast rotators); also see \Fig{fig:vtheta}.
%%the meridional flow with the bulk is more, but the surface flow is the least, which again gives a weak polar field.
%%And as we decrease the rotation period from this point, then the surface flow increases and again gives more polar field. Hence we observe a increase in polar field from both sides while coming from low meridional bulk flow to high or from low surface flow to high surface flow.
%The time-latitude plot, as shown in  \Fig{fig:stel_bfly}, also clearly shows the effect of meridional flow in different ranges of stars.
%}

%%As the meridional flow becomes stronger with the increase of stellar rotation period in \citet{KO11} model, in \Fig{fig:stel_dyn}, the meridional flow becomes stronger along the horizontal axis; see the top axis of the panel (b) where the average value of $v_\theta$ over $r/R_{\rm s} = 0.9$ -- 1 is shown. 
%%Therefore, the trend in \Fig{fig:stel_dyn}(a) should be 
%%similar to the ones obtained for variations of poloidal and toroidal fields with meridional circulation speed (\Figs{fig:omeff}{fig:dyn}).
The decline trend of magnetic field as observed in \Fig{fig:stel_dyn}(a) beyond about a ten-day rotation period gives one contribution of the observed decrease of the stellar cycle strength with the rotation period \citep{Noyes84a,BoroSaikia18}. 
The decreasing trend at shorter rotation periods also helps to explain the saturation of the magnetic field in rapidly rotating stars. 
This is because in reality, for the rapidly rotating stars,  the strength of the poloidal field generation (\bl\ process and $\alpha$ effect) should be stronger \citep{KO15}, and the diffusivity should be weaker \citep{KPR94, Kar14b}. These two will tend to increase the magnetic field, and thus, the resultant effect will be a magnetic field saturation in the very rapidly rotating stars.

Finally, the cycle period shows an interesting trend, as shown in \Fig{fig:stel_dyn}(b). 
This variation can be explained from the variation of $v_\theta(R_s, 45^\circ)$ and $\overline{v}_\theta^{\rm pole}$ at $45^{\circ}$ latitude as presented in \Fig{fig:vtheta}. The initial increase of cycle period with the rotation period is due to the weakening of surface flow, while the decline of cycle duration beyond about 7 days rotation period is due to the strengthening of bulk poleward flow. 
The effect of the strong (weak) surface flow on the poleward transport of the radial magnetic field is clearly depicted in the butterfly diagram for the 1 day (7 and 25.38 (solar value) days) rotation period in \Fig{fig:stel_bfly} 

As for rapidly rotating stars, the cycle period vs rotation period trend is less obvious in observations \citep{BoroSaikia18}; we cannot make a comparison with observations. We find that for slowly rotating stars, we are getting the opposite trend. This decreasing trend of the cycle period at a large rotation period is due to the enhancement of the meridional flow. Thus, we need additional physics to explain this observed trend \citep{KTV20, Hazra19, V23, K22}.

\section{Conclusions} \label{sec:conclusion}
In this comprehensive study, we revisited the pivotal role of meridional flow in generating magnetic fields in the Sun and sun-like stars, using dynamo models.
%Using dynamo models, our comprehensive study revisited the role of meridional flow in generating magnetic fields in the Sun and sun-like stars. 
We show that the STABLE dynamo model, which captures the \bl\ process through deposition and decay of BMRs on the solar surface, represents the generation of the poloidal field in the Sun more realistically than the dynamo models with explicit $\alpha$ effect parameterization for the \bl\ process. In particular, the variation of the polar field strength with the meridional flow matches the variation found robustly in SFT models \citep[e.g.,][]{Bau04}. 
Specifically, the polar field strength increases with a moderate increase in meridional flow speed but decreases once the flow exceeds a certain value.
When the meridional flow is negligible, the cross-equatorial cancellation is poor. With the increase in flow speed, the trailing polarity flux is dragged efficiently by the flow, increasing the polar field strength. A further increase of meridional flow causes both leading and trailing polarity flux to move to the pole, which decreases the polar field. 

Moreover, our analysis extends to the toroidal field, which exhibits a similar trend to the polar field: increasing at moderate meridional flow speeds and then decreasing. 
This behavior is due to two reasons. 
One is the variation of the polar field itself with the flow, and the other is due to the competition between the shearing and diffusion effects. When meridional flow is small, the magnetic fields stay in the CZ for a longer time, giving more time for the diffusion of the fields. Thus, increasing meridional flow increases the toroidal field. However, when the meridional flow is too strong, the shear gets very little time to induce a toroidal field, thereby decreasing the toroidal field with the further increase of flow speed. 
Eventually, the strength of the magnetic cycle shows an initial increase followed by a decreasing trend with the meridional flow speed.

Furthermore, we observe similar trends for both poloidal and toroidal fields in relation to meridional flow in dynamo model. 
%Next, we show similar poloidal and toroidal fields vs. meridional flow trends in dynamo mode. 
This behavior is attributed to the combined effect of meridional flow on poloidal field generation and the diffusion of the magnetic field, as explained above. 

Most importantly, our study identifies the role of meridional flow in determining the cycle strength and duration of stellar cycles. In solar-like stars, the meridional flow speed is expected to vary with the rotation rate of the stars \citep{Brown08, viv18}. Mean-field model of \citet{KO12b} suggests that the rapid rotators typically exhibit higher velocities near the surface but lower flow in the bulk of the convection zone, whereas slow rotators possess stronger flows in the bulk. By including the meridional circulation data from the model of \citet{KO12b} in our STABLE dynamo model, we show that the strength of the magnetic field first increases with the stellar rotation rate and then declines at rapid rotation. The increasing trend suggests that the meridional flow is one of the contributors to the enhancement of magnetic activity with the increase of the rotation rate of solar-like stars. The decline of the magnetic field strength at rapid rotation can help to compensate for the increase of field due to the enrichment of dynamo efficiency at rapid rotators and thus provides an explanation of the the saturation of magnetic activity in rapid rotators. 
Therefore, the variation of meridional flow with stellar rotation is an important component for explaining the features of stellar cycles at different rotation rates.

\section*{Acknowledgement}
We would like to acknowledge the contributions of Leonid Kitchatinov for
providing the data of differential rotation and meridional flow from \citet{KO12b} model for different stars, for carefully checking the manuscript, providing valuable feedback, and raising insightful questions, which enhanced the quality of the paper.
We also wish to acknowledge the anonymous referee for their thorough review of the manuscript and for their insightful comments which helped in improving the manuscript. V.V. acknowledges the financial support from the DST through the INSPIRE fellowship. B.B.K. acknowledges the Science and Engineering Research Board (SERB) for providing financial support through the MATRIC program (file no MTR/2023/000670).  
The computational support and the resources provided by the PARAM SHIVAY Facility under the National Supercomputing Mission, the Government of India, at IIT (BHU) Varanasi, is gratefully acknowledged.

\section*{Appendix}

\begin{figure}[h]
\centering
\includegraphics[width=\columnwidth]{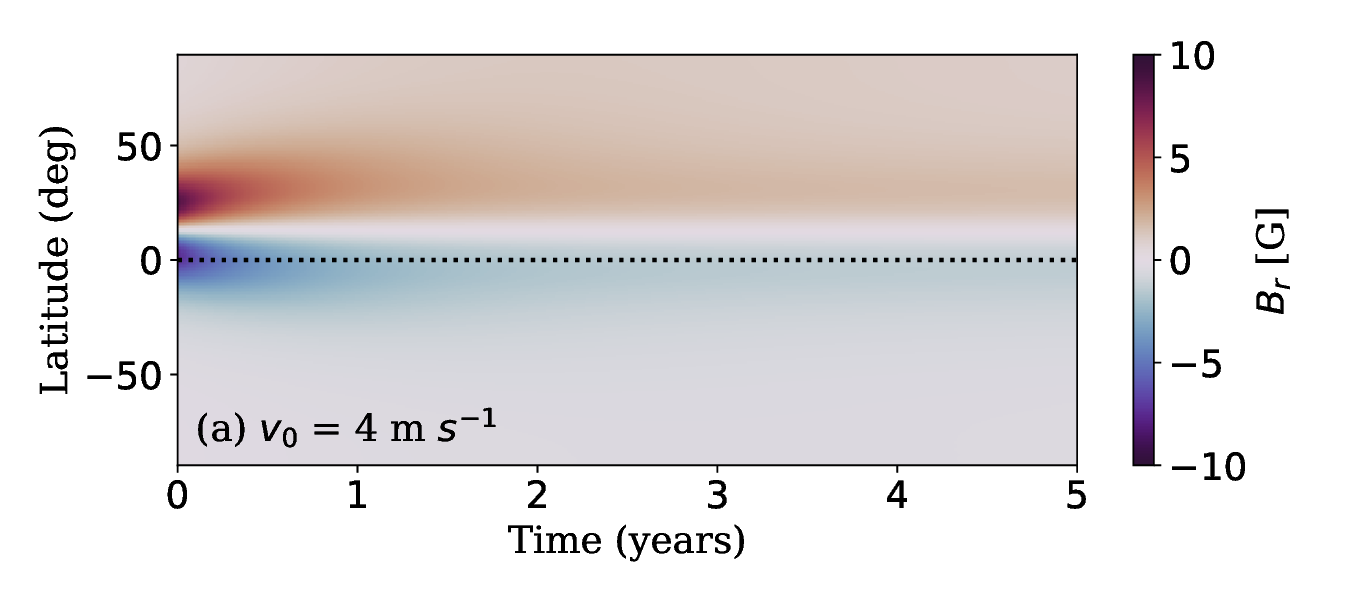}
\includegraphics[width=\columnwidth]{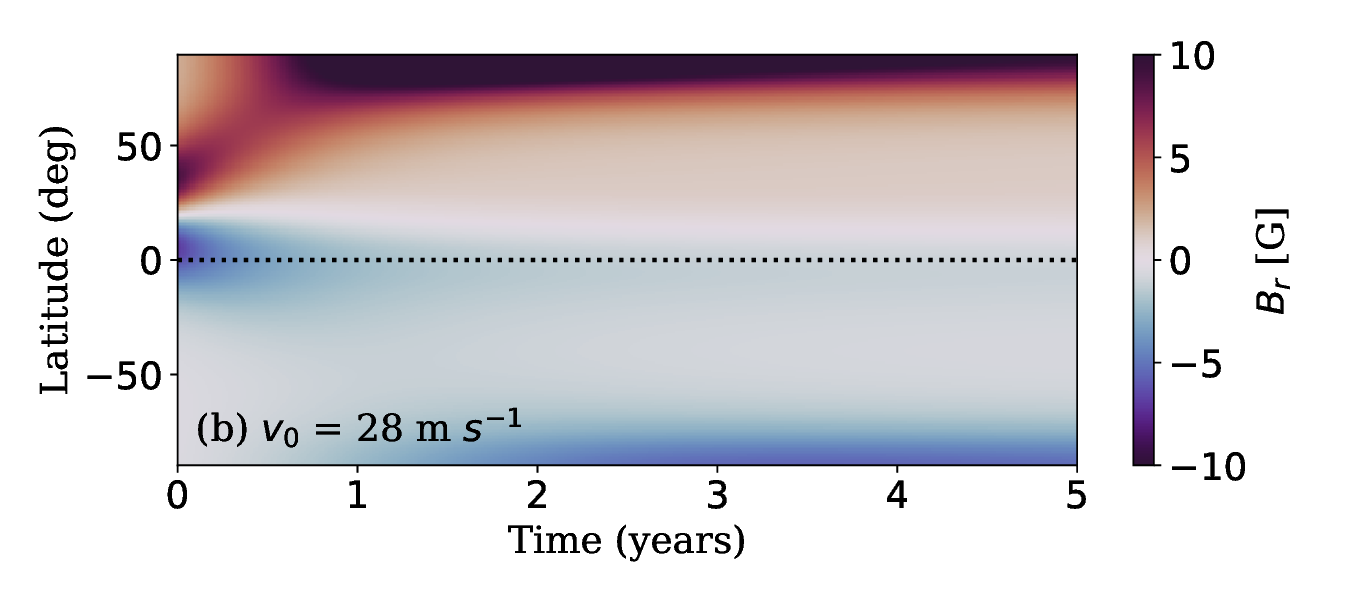}
\label{fig:bfly2d}
\caption{Time-latitude plot of the radial field from the 2D model with \bl\ $\alpha$ for (a) $v_0$ = 4 \mps\ and (b) $v_0$ = 28 \mps.}

\end{figure}

\begin{figure}
\centering
\includegraphics[width=\columnwidth]{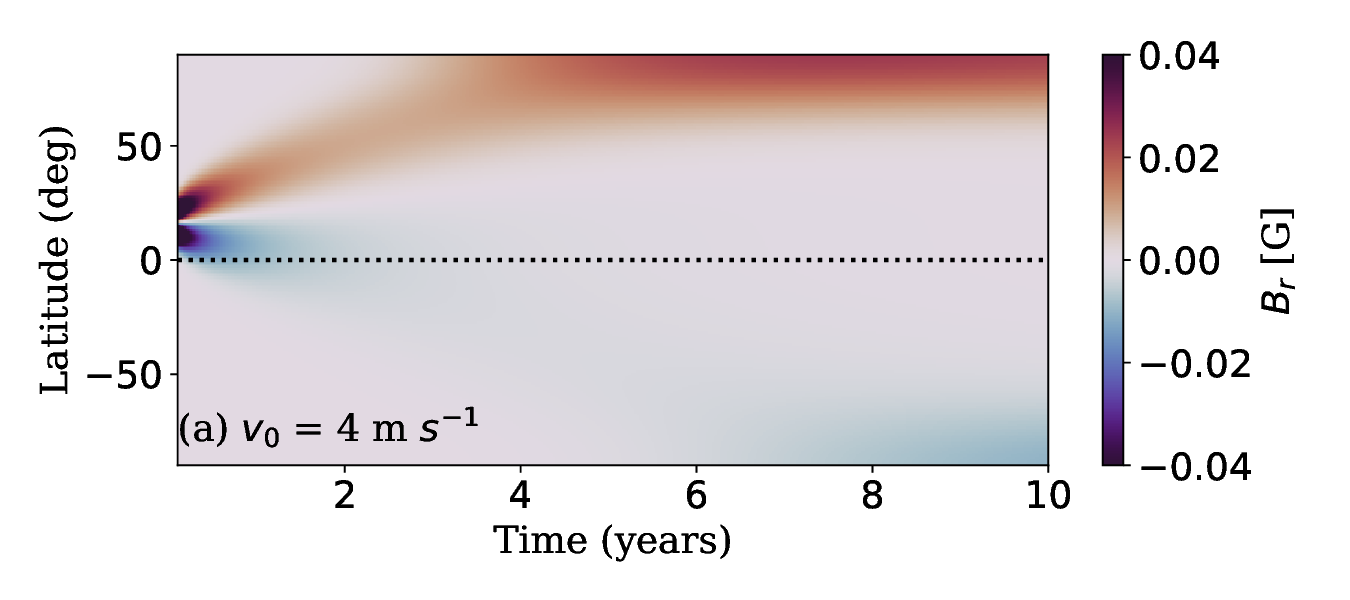}
\includegraphics[width=\columnwidth]{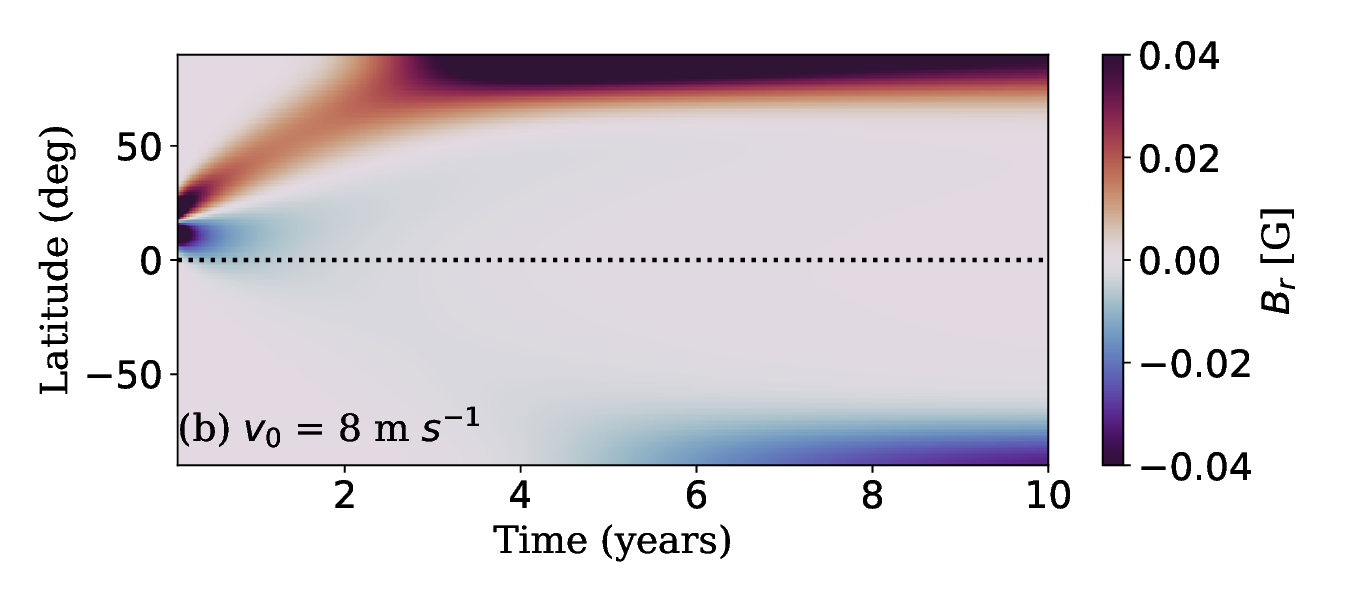}
\includegraphics[width=\columnwidth]{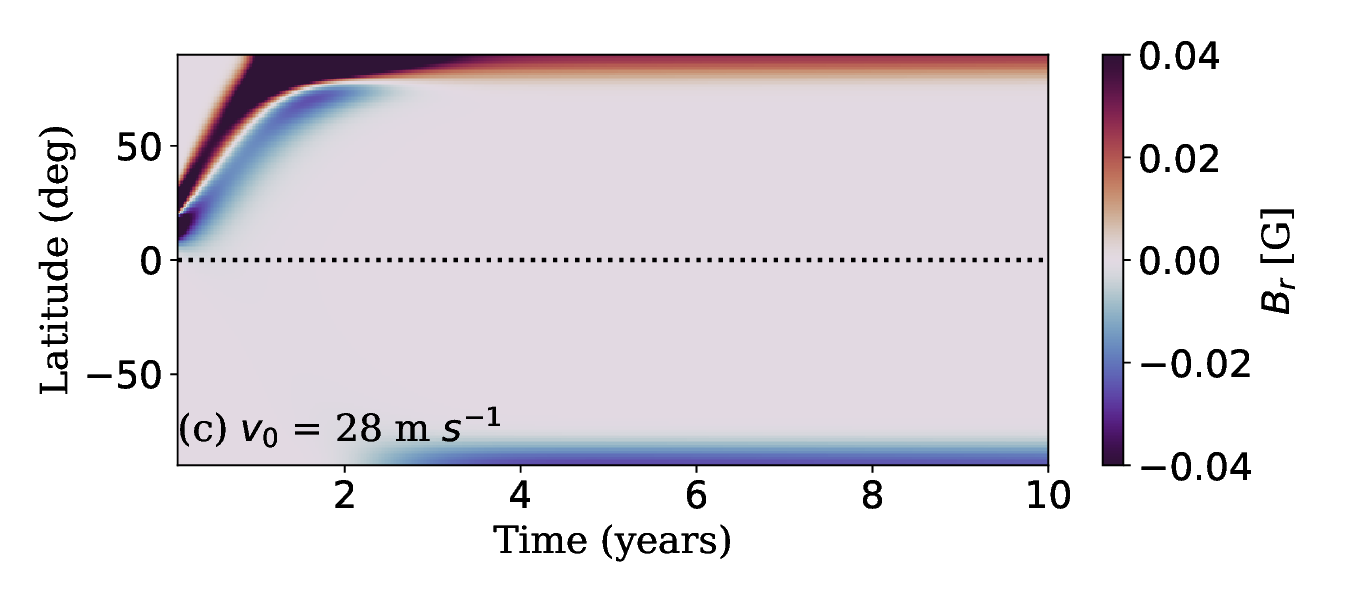}
\label{fig:bfly3d}
\caption{Time-latitude plot of the radial field from the 3D STABLE model with a single BMR for (a) $v_0$ = 4~\mps, (b) $v_0$ = 8~\mps, and (c) $v_0$ = 28 \mps.}

\end{figure}

\newpage
\bibliography{paper}{}
\bibliographystyle{aasjournal}



\end{document}